\documentclass[12pt]{article}
\usepackage{graphicx}
\usepackage{amsmath}

\newcommand{\bb}{{\bf b}}

\newcommand{\bx}{{\bf x}}
\newcommand{\by}{{\bf y}}

\newcommand{\bbeta}{\mbox{\boldmath $\beta$}}

\newcommand{\bfgamma}{\mbox{\boldmath $\gamma$}}

\newcommand{\bsigma}{{\bf \Sigma}}

\begin{document}

\title{An ensemble approach to short-term forecast of COVID-19 intensive care occupancy in Italian Regions}

%\titlerunning{Short-term forecast of Covid-19 ICU occupancy in Italy}        % if too long for running head

\author{Alessio Farcomeni \\ \small Dipartimento di Economia e Finanza, \\ \small Universit\`a di Roma ``Tor Vergata'' \\ \\ Antonello Maruotti \\ \small Dipartimento GEPLI\\ \small Libera Universit\`a Maria Ss Assunta\\ \small Department of Mathematics\\ \small University of Bergen\\ \\ \\
        Fabio Divino \\ \small Dipartimento di Bioscienze e Territorio\\ \small Universit\`a del Molise \\  \\ Giovanna Jona Lasinio \\ \small Dipartimento di Scienze Statistiche\\ \small Sapienza Universit\`a di Roma\\ \\ Gianfranco Lovison  \\ \small Dipartimento di Scienze Economiche, Aziendali e Statistiche\\ \small Universit\`a di Palermo \\ \small
Department of Epidemiology and Public Health\\ \small Swiss TPH Basel \\
}

%\institute{
%A. Farcomeni \at
%
%\email{alessio.farcomeni@uniroma2.it}           
%\and
%A. Maruotti
%\at
%Dipartimento GEPLI, Libera Universit\`a Maria Ss Assunta - Roma\\
%Department of Mathematics, University of Bergen\\
%\email{a.maruotti@lumsa.it}
%\and
%F. Divino \at
%Divisione di Fisica, Informatica e Matematica - Dipartimento di Bioscienze e Territorio, Universit\`a del Molise - Pesche(IS) \\
%\email{fabio.divino@unimol.it}
%\and
%G. Jona Lasinio \at
%Dipartimento di Scienze Statistiche, Sapienza Universit\`a di Roma \\
%\email{giovanna.jonalasinio@uniroma1.it}
%\and
%G. Lovison \at
%Dipartimento di Scienze Economiche, Aziendali e Statistiche, Universit\`a di Palermo \\
%Department of Epidemiology and Public Health, Swiss TPH Basel \\
%\email{gianfranco.lovison@unipa.it}
%}  

\date{}

\maketitle

\begin{abstract}
 The availability of intensive care beds during the Covid-19 epidemic is crucial to guarantee the best possible treatment to severely affected patients.
  In this work we show a simple strategy for short-term prediction of Covid-19 ICU beds, that has proved very effective during the Italian outbreak in February to May 2020.
  Our approach is based on an optimal ensemble of two simple methods: a generalized linear mixed regression model which pools information over different areas,
  and an area-specific non-stationary integer autoregressive methodology. Optimal weights are estimated using a leave-last-out rationale.
  The approach has been set up and validated during the epidemic in Italy. A report of its performance for predicting ICU occupancy at Regional level is included.\\
{\bf keywords: }{Integer autoregressive model \and Generalized Linear Mixed model \and Clustered data \and Weighted ensemble \and SARS-CoV-2 \and Covid-19}
\end{abstract}

\section{Introduction}

Italy has been, and partially still is, under pressure to properly manage the recent COVID-19 epidemic emerged from China in December 2019. Its quick spread required a global response to prepare health systems worldwide. In its present form, COVID-19 seems to have two very challenging characteristics (see also \cite{Peeri2020}): it is highly infectious and, despite having a benign course in the vast majority of patients, it requires hospital admission and even intensive care for a far from negligible proportion of infected.

In Italy, particularly in the two Regions of Lombardia and Veneto, the COVID-19 infection emerged in February 2020 with a basic reproductive number $R_0$ between 3 and 4 \cite{flaxal:2020}. At the beginning of the coronavirus outbreak, Italy was one of the countries with the lowest amount of acute care beds per person in Europe.  The resources of the national health system were not designed to face a large-scale epidemic. The national health system was equipped with a total number of approximately 5200 beds, that was substantially increased very quickly   during the spread of the disease \cite{remuzzi2020}. This reversed a trend that started years ago, especially after financial crises, according to which resources allocated to the national health system were progressively cut.

In general, ICUs are characterized by a rather low number of beds with high turnover.
Most of patients stay in intensive care only for one or few days \cite{diekal:2013}, even if some of them can stay months.
However, ICU length of stay (LOS) due to COVID-19 infection was rather long: a recent study \cite{grasal:2020} involving 1591 patients admitted in ICUs in Lombardia reported a
median length of stay of 9 days (6-13 [95\% CI]). Long LOS implied a slower turnover, and increased the risk of collapse of the national ICU system.
Indeed, in the absence of measures to flatten the epidemic curve, the number of ICU beds available in Italy would have achieved very quickly 100\% occupation,
leaving several thousand patients short of the cures needed (see also \cite{remuzzi2020} for a discussion).

It is now clear that measures like social distancing, use of masks, contact tracing, and a wide access to diagnostic testing coupled with isolation of positives
can be very effective. Nevertheless, careful and reliable planning of resources can also aid
substantially in controlling the consequences of the epidemics, and likely increase the likelihood of early diagnoses and better care.
To respond to the looming threat of shortage of ICU beds, hospitals urgently need to establish and implement policies that more fairly allocate these scarce resources.
If hospitals can plan in advance how many ICU beds shall be made available for the nearly following days, capacity can be increased (or decreased) to match the demand.
This would avoid the ethical dilemma of severe triaging patients and not admitting those whose lives are {\it not worth saving} \cite{white2020}.
In this work we propose one statistical tool of this sort, which can be used to accurately forecast the ICU occupation for the next one to five days.
Beds demand in intensive care depends on two factors: the number of COVID-19 patients needing intensive care, and duration of their hospitalisation. Unfortunately,
these data are not made available during the Italian outbreak. Public data include, anyway, daily ICU occupation by Region, which we use as a proof of concept.

Existing approaches to forecast ICU occupancy are mainly based on exponential models fitted to daily numbers of occupied critical care beds out
of confirmed cases \cite{grasselli2020b,sebastiani2020} or on  SIR (susceptible, infectious, recovered) models \cite{Giordano2020}.
Both approaches are subject to certain limitations in our opinion.
An obvious limitation of the former approach includes ascertainment bias due to the use of counts of confirmed cases, and inappropriate modeling assumptions (including for instance
a Gaussian assumption for log-counts).
A model of the SIR family may on the other hand be a very appropriate option. However, SIR-based models require the possibility to precisely estimate several
characteristics of the epidemics which are still mostly unknown, and it is well known that changing even slightly the initial conditions can lead to very different
results. This is even more difficult as the exposed population is only partially observed \cite{Bohning2020,Yue2020,Chen2020}.

Our approach is based on optimally combining two forecasting methods. The first is based on Poisson mixed effects regression; the second one is a Region-specific time series model for counts, taking into account time-depen\-den\-ce over time. The count outcome is appropriately modeled as a Poisson conditionally on observed time trends and unobserved heterogeneity including dependence, as implied by random effects or by the auto-regressive structure of the time-series models, and the averaged predictions give an optimal balance between pooling information over different areas (which targets a low variance prediction) and adaptation at the specific area (which targets a low bias prediction).

The rest of the paper is as follows: in the next section we give a description of the available data. In Section \ref{method} we describe our method for
prediction of the next-day ICU occupancy. We mention here that we have validated our method
during the outbreak, by repeatedly producing ICU predictions with the current data, and waiting for the official data for the next five days.
An illustration of the performance of our approach is given in Section \ref{results}. Some concluding remarks are in Section \ref{conclusions}.

An implementation of our approach for the Italian case is available within our Shiny app at: {\tt https://statgroup19.shinyapps.io/StatGroup19-Eng}.

\section{Data}
\label{data}

\begin{figure}
\centering
\includegraphics[scale=0.5]{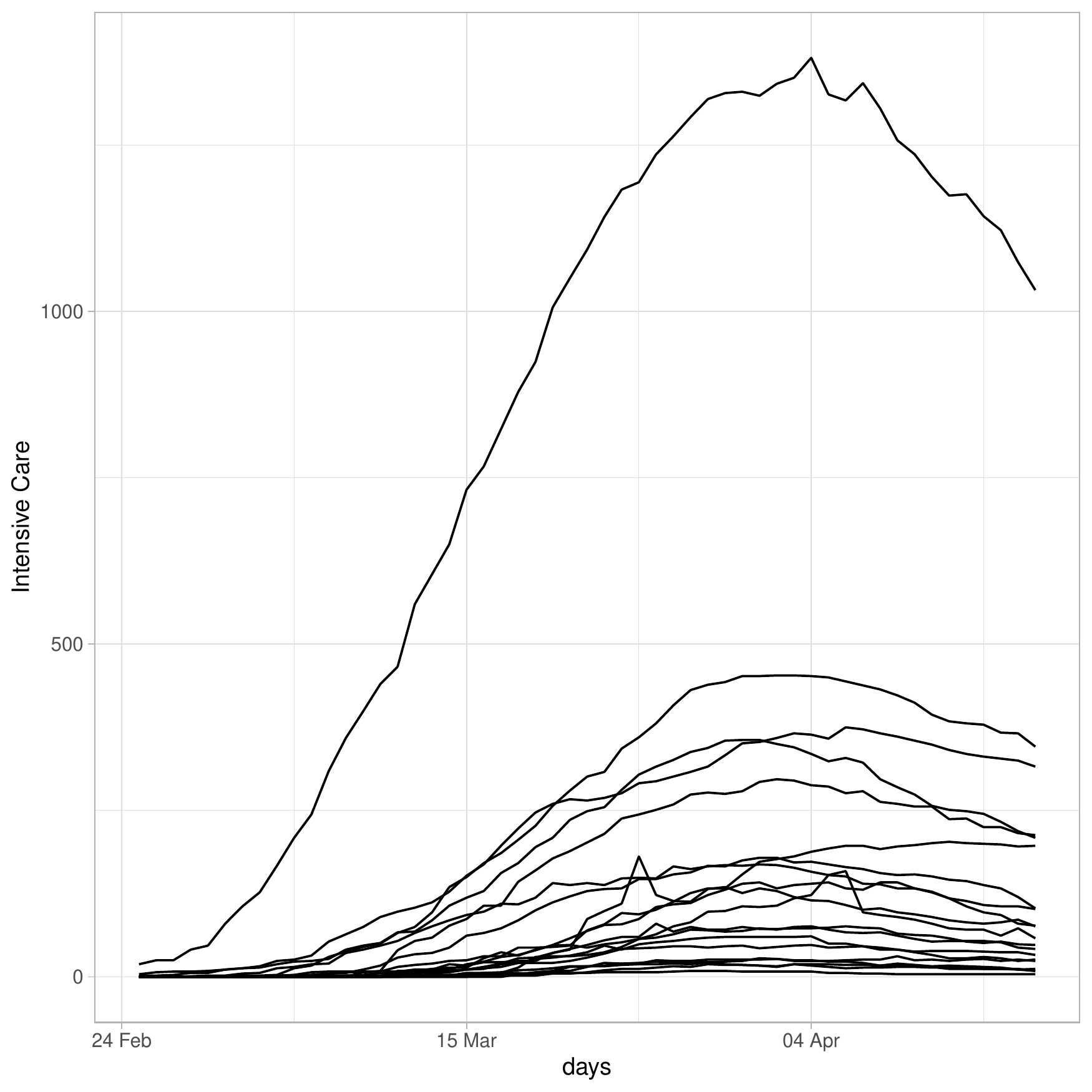}
\caption{Regional time series of ICU admission from February 24, 2020 until April 16, 2020}\label{fig:longitudinal}
\end{figure}

\begin{figure}
\centering
\includegraphics[scale=0.5]{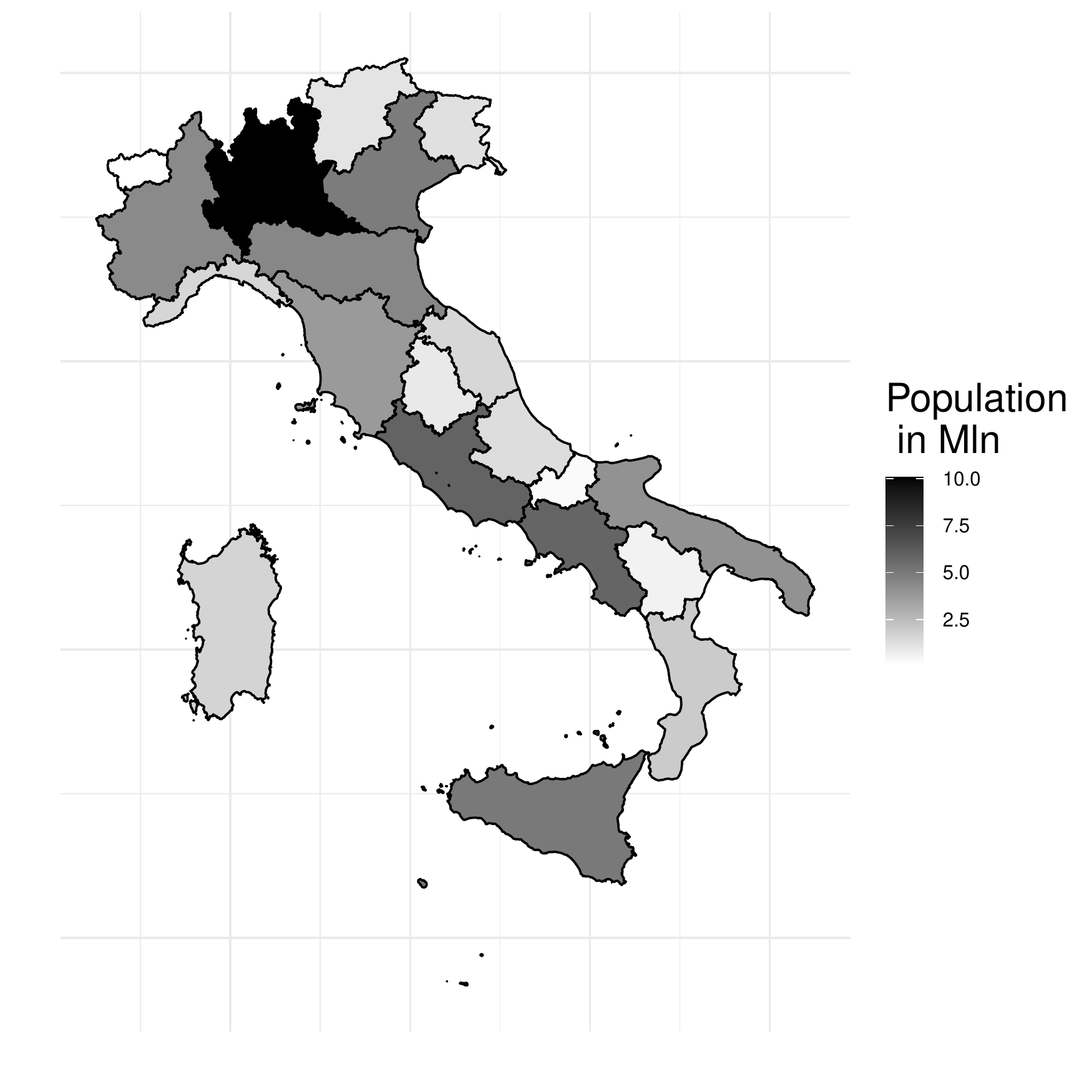}
\caption{Italian Regional population size}\label{fig:popmap}
\end{figure}

On January 31, 2020, the Italian Government declared the state of emergency for six months as a consequence of the health risk associated with COVID-19 infection.\\
To inform citizens and make the collected data available, the Department of Civil Protection  developed an interactive geographic dashboard (accessible at the addresses http://arcg.is/C1unv (desktop) and http : //arcg.is/081a51 (mobile)) and built a daily updated data github-repository (CC-BY-4.0 license), with a large number of variables; among them the intensive care defined as \emph{number of hospitalized patients in Intensive Care Units} (ICU) available at the Regional level (20 Italian Regions). In Figure \ref{fig:longitudinal} the time series of ICU admission from February 24 until April 16 , 2020 are reported and it appears that several sources of heterogeneity affect the data. The epidemic started at different times across Regions, starting from the North of the country and evolved very differently in each Region, leading to trajectories of different shapes. Population size is  also very different across Italian Regions (Figure \ref{fig:popmap}), which are also characterized by different economic and social structures. Moreover, a further heterogeneity source relates to the ICU capacity.
In Italy, the health system is Regional-based, so in practice there are 20 different health systems and management choices, local authorities and local policy makers play a fundamental role in the definition of health services capacities, with very few national directives.

\begin{figure}
\centering
\includegraphics[scale=0.5]{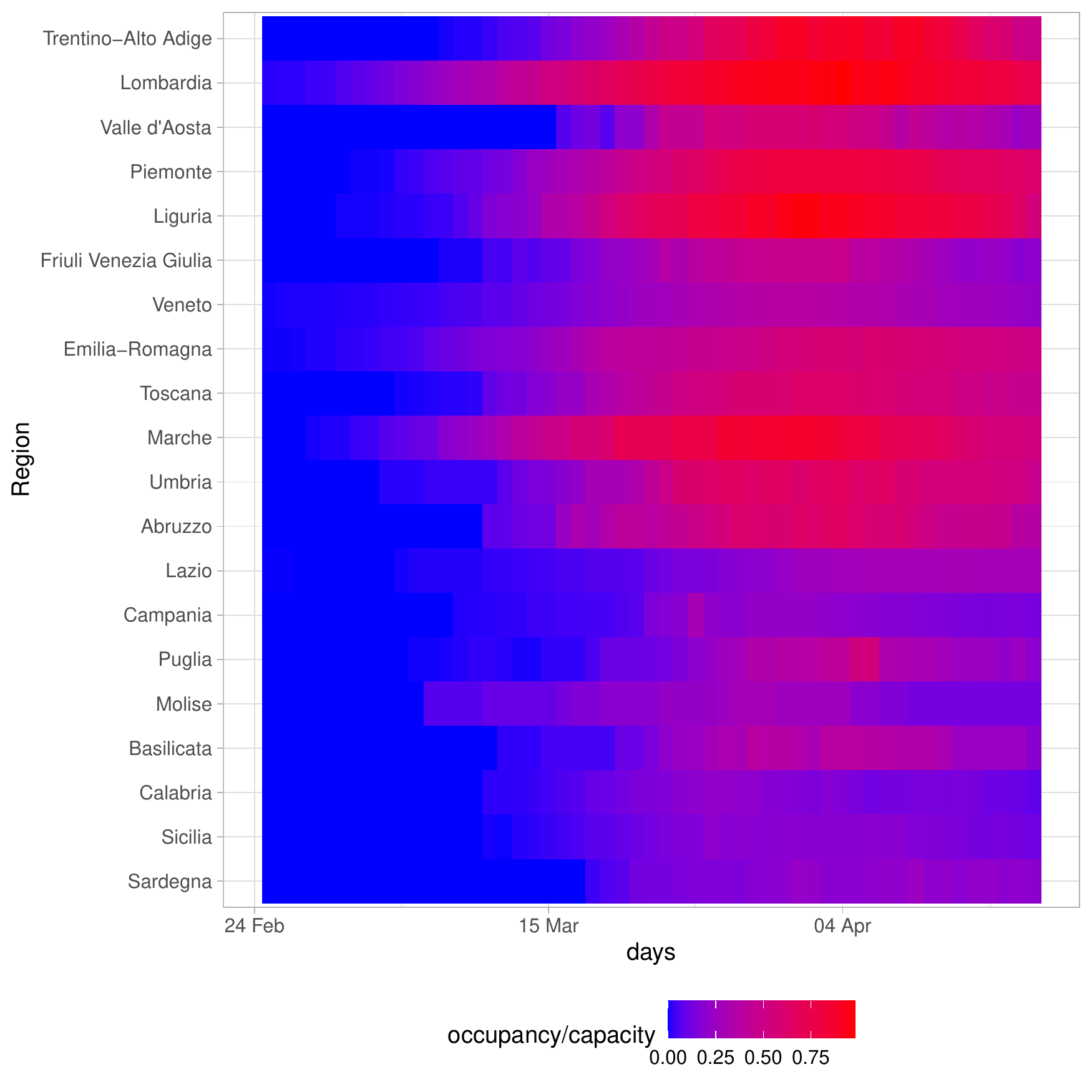}
\caption{ICU rate of occupancy during the COVID-19 epidemic in the Italian Regions. Regions are ordered geographically from North to South.}\label{fig:heatmap}
\end{figure}

During the epidemic some of the northern Regions went beyond the capacity of the local system. Figure \ref{fig:heatmap} reports the Regional rate of occupancy (occupancy over capacity\footnote{Notice that in figure \ref{fig:heatmap} we use the capacity on April 16, 2020 that was augmented in the northern Regions to face the COVID-19 emergency}) during the epidemic. Regions are ordered geographically from North to South showing the north-south gradient of the epidemic evolution mostly affecting the northern Regions. It is rather clear that the health systems of northern Regions were under pressure for a longer period and this reflects the difference in the spread of the epidemic. A Regional model that takes into account different sources of heterogeneity and the different timing in the spread of the epidemic is thus required.

\section{Methods}
\label{method}

In order to produce accurate forecasts we combined two methods, one based on joint modelling all Regions through a mixed-effects generalized linear regression model,
and the other one based on separately modelling each Region as a non-stationary time-series of counts, which uses an integer-valued autoregression specification, with covariates.
The first approach pools information over Regions, reducing variance of the final prediction; the second instead can be expected to have a lower bias. The final predictions are then averaged with weights chosen through a leave-last-out method. Furthermore, in order to focus more on the recent trends, we always set $T=15$, that is, we use the last two weeks of data to make predictions and ignore previous measurements.

\subsection{Random effects modelling of longitudinal count data}

We start assuming that the observed daily ICU admissions for Region $i$ at day $t$, $y_{it}$, are realizations of independent
Poisson random variables $Y_{it}$ with parameter $\mu_{it}$, $\forall i = 1,\dots,I, \; t=1,\dots,T$.
The interest is in modelling $\mu_{it}$ as a non-linear function of time, taking into account unobserved heterogeneity and Region-specific effects. A simple way to deal
with these features is through a Generalised Linear Mixed Model (GLMM) \cite{breslow1993} with linear predictor

\begin{equation}\label{eq:glmm}
\log(\mu_{it}) = (\beta_{0}+b_{0i}) + (\beta_{1}+b_{1i})\times t +(\beta_{2}+b_{2i})\times t^2 + \log(residents_i)
\end{equation}

where a canonical link has been adopted, the offset term $\log(residents_i)$ accounts for different population exposures, $\boldsymbol{\beta} = (\beta_0,\beta_1,\beta_2)$ represents the vector of shared fixed-effects regression parameters, and ${\bf b}_i = (b_{0i},b_{1i},b_{2i})$ represents the random coefficients, i.e. the Region-specific intercept and slopes, with $${\bf b}_i = N\left({\bf 0}, \bsigma_B \right)$$

The observed counts are assumed independent given the three-dimensional random vector ${\bf b}_i$. The likelihood function is obtained integrating out the ${\bf b}_i$ in \eqref{eq:like.glmm}

\begin{equation}\label{eq:like.glmm}
L(\bbeta, \bsigma_B; \by) = \prod_{i=1}^I \left\{\int_{\Re^3} \prod_{t=1}^T \frac{\mu_{it}}{y_{it}!} \exp(- \mu_{it})  \frac{1}{\mid 2 \pi \bsigma_B \mid^{\frac{1}{2}}}
           \exp \left( - \frac{1}{2} \bb_i^T \bsigma_B^{-1} \bb_i \right) d \bb_i \right\}
\end{equation}

As widely discussed in the literature, the integral in (\ref{eq:like.glmm}) has no analytical solution, and approximate methods should be used in order to estimate the parameters  $\bbeta$ and $\bsigma_B$. Here, a Laplace approximation has been used first of all to speed up computations. Adaptive Gaussian quadrature may represent a slightly more accurate option, at the expense of a higher computational burden. Other approaches can also be considered \cite{zhang2011,kim2013}.

Predictions are based on the posterior estimates of the random effects and the MLE of the fixed-effect parameters. Predictions intervals are found through non-parametric block bootstrap using 500 replicates. Block bootstrap involves resampling Regions, and once a Region is included its entire time-series is used for model estimation of the resampled data.
This is done to preserve the dependency structure of the data.

It is worth mentioning that the specific covariance structure among the random effects has been chosen
on the grounds of model selection using BIC. In our data, the best covariance structure, which has been
then used for all estimates and predictions, has turned out to be:
\begin{equation}
\bsigma_B =
\begin{bmatrix}\sigma^2_0 & \sigma_{01} & 0 \\
\sigma_{01} & \sigma^2_1 & 0\\
0 & 0& \sigma^2_2\end{bmatrix}
\end{equation}

\subsection{Region-specific integer-valued autoregression modelling}

At the second step, we fit and obtain predictions for Regional time series separately.
In other words, 20 different models are fitted, as time-series models for counts.

Let again $Y_{it}$ be the daily number of ICU admissions and let ${\bx}_{it} = (t^0,t^1, \ldots,$ $t^r)^T$ denote
an ($r+1$)-dimensional time-varying covariate vector, consisting of a polynomial specification of time; notice that this
vector is Region-specific, so different polinomial specifications can be selected for different Regions.
We model $Y_{it}$ as a conditional Poisson distribution where the expectation $\mu_{it}$ at time $t$ depends
on both past counts and past covariates:
\begin{equation}
\label{eq:mslinear}
%\lambda_t = a \lambda_{t-1} + b Y_{t-1}  + \exp\left(\alpha +  \beta'{\bf X}_{t-1}\right) , \quad t\geq 1.
\mu_{it} = \alpha_0 \mu_{i,t-1} + \alpha_1 Y_{i,t-1}  + \bfgamma^T {\bx}_{i,t-1} , \quad t>1.
\end{equation}
where the coefficients $\alpha_0$ and $\alpha_1$ represent the effecst of the expectation $\mu_{i,t-1}$ in the previous day
and the number of ICU admissions in the previous day $Y_{i,t-1}$, respectively.

Note that the model defined by \eqref{eq:mslinear}  belongs to the INGARCHX family
\cite{fokianos2009,agosto2016,chen2016}. An alternative approach would be to examine the log-linear
INGARCH model of \cite{fokianos2011}. However, the linear model has significant advantages in terms
of interpretation, since it allows for an additive decomposition of the expectation.
Under the Poisson assumption, equations (\ref{eq:mslinear}) corresponds to a GARCH-type model
\cite{bollerslev1986} for the conditional variance of the process; hence the name INGARCH
has been used frequently in the literature (though there is some debate on this terminology, 
see e.g.\cite{tjostheim2012}).

For each Region, we compare stationary, linear, quadratic and cubic trends (that is, $r=0,1,2,3$).
We select the best model specification for each one separately,
according to the Bayesian information criterion.

Parameters in equation (\ref{eq:mslinear}) are estimated via conditional maximum quasi-likelihood estimation,
using the function {\tt tsglm} in the \textbf{tscount} {\tt R} package.
If the Poisson assumption holds true, then we obtain an ordinary ML estimator. Predictions and their confidence 
intervals are obtained as follows. The distribution of the
1-step-ahead prediction is not known analytically and it is thus approximated numerically
by a parametric bootstrap procedure. One-step-ahead prediction intervals can be straightforwardly
obtained and are based on simulations of realizations from the fitted model; the approximated
prediction intervals are obtained from the empirical $2.5\%$ and $97.5\%$ quantiles of the boostrap-based predictions.

\subsection{Model averaging}

Let $\hat y_{i,T+1}^{(1)}$ denote the prediction obtained for the $i$-th Region ($i=1,\ldots,20$ for the Italian case) and time $T+1$
with the GLMM method, and $\hat y_{i,T+1}^{(2)}$ the prediction obtained with the integer autoregressive method. The final prediction is

\begin{equation}
\hat y_{i,T+1} = w_{i,T+1} \hat y_{i,T+1}^{(1)} + (1-w_{i,T+1}) \hat y_{i,T+1}^{(2)},
\end{equation}

for some $w_{i,T+1} \in (0,1)$. In order to estimate an optimal $w_{i,T+1}$, we first repeat model estimation
excluding $y_{iT}$ for $i=1,\ldots,I$; obtaining leave-last-out predictions
$\hat y_{iT}^{(1)}$ and $\hat y_{iT}^{(2)}$; and then we solve the optimization problem
$$
w_{i,T+1} = \arg\inf_{x \in (0,1)} \left| x \hat y_{iT}^{(1)} + (1-x) \hat y_{iT}^{(2)} \right|.
$$
It shall be here noted that for the first few days, when $T<15$, we make the weight-homogeneity
assumption $w_{i,T+1}=w_{j,T+1}$ for all $i\neq j$.
For reasons of simplicity final prediction intervals are obtained as the weighted average of the limits of
prediction intervals for $\hat  y_{i,T+1}^{(1)}$ and $\hat y_{i,T+1}^{(2)}$.
It is straightforward to use Jensen's inequality to show that this conservatively guarantees the nominal level.

\section{Results}
\label{results}

The reliability and goodness of our approach can be assessed by checking the next-day performance as:
(i) median absolute error over the twenty Italian Regions, (ii) mean relative error over the twenty Italian Regions,
(iii) proportion of prediction intervals that do not contain the actually observed occupancy,
(iv) proportion of observed occupacies above the upper limit of the prediction interval.
For illustrative purposes we discuss in this section predictions related to the next day, for days from
March, 17 to April, 27th.

\begin{figure}[!ht]
   \centering
 \begin{center}
  \includegraphics[height=8cm,width=9cm]{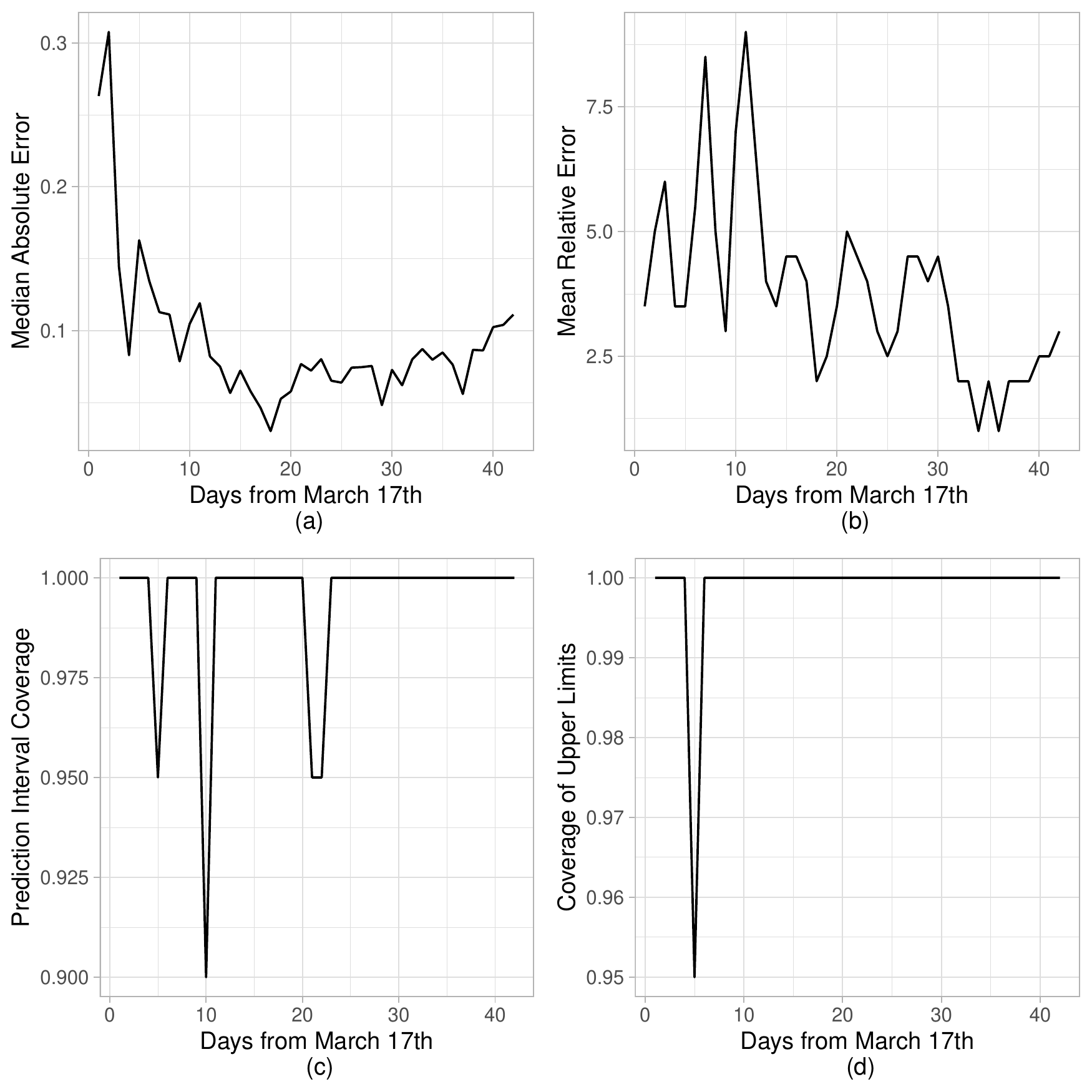}
  \end{center}
 \caption{Daily median absolute prediction error (a), daily mean relative prediction error (b), daily coverage of prediction intervals (c)
   and daily coverage of upper limits of prediction intervals (d) for our ensemble prediction method for ICU occupation in the twenty Italian Regions.}
		\label{plotResults}
\end{figure}

The daily absolute error has a median of 4 beds, with first quartile 1 and third quartile 8.
The daily relative error over the twenty Regions has first quartile 2\%, median 5\%, third quartile 12\%.
Its mean is 9.2\%. For prediction intervals we used a nominal level of 99\%. Out of the 840
intervals produced, 99.4\% indeed contained the observed ICU occupation.
Figure \ref{plotResults} summarizes the main results. We report a plot of the next-day performance for each day since March, 17. It can be seen that after the first two weeks, when data available were scarce and a unique weight was used, the absolute and relative errors decreased substantially. Furthermore, coverage of the prediction intervals was always very close to 100\%, with only one occasion in which one Region reported an ICU occupation above the predicted upper limit. Overall, the approach is rather effective and needs just few data points to produce reasonable estimates.

Just to corroborate the results, we provide an example of the forecasts shown daily at {\tt https://statgroup19.shinyapps.io/StatGroup19-Eng}. On April 9$^{th}$, we published the forecasts displayed in Table \ref{tab:icu} on the web, along with 99\% confidence intervals. On April 10$^{th}$, we checked how well the ensemble approach performs. Overall, the performance of the proposal is more than satisfactory. The observed values are all into the 99\% confidence intervals, whose length is reasonable and ensure a good representation of the uncertainty surrounding the forecasts. The upper bound of the confidence interval can be used as the worst possible scenario for a specific day and resources should be allocated accordingly to guarantee optimal health services. To be fair, the confidence interval for Puglia is rather wide. However, this is not surprising, as patients were transferred from northern Regions to Puglia for a few days, and daily bursts in the time series of ICU admissions were observed.

\begin{table}[htb]
\begin{center}
\caption{Forecasted and observed ICU beds at the regional level: April 10$^{th}$.\label{tab:icu}}
\begin{tabular}{lccc}
\hline
Regions &  Observed & Forecast & Confidence Interval ($99\%$)\\
\hline
Abruzzo & 53 & 57 & 48 -- 91\\
Basilicata & 15 & 17 & 8 -- 29\\
Calabria & 14 & 13 & 5 -- 23 \\
Campania & 90 & 87 & 60 -- 107\\
Emilia-Romagna &  349 & 347 & 301 -- 397\\
Friuli Venezia Giulia & 33 & 34 & 19 -- 48\\
Lazio & 201 & 197 & 155 -- 225\\
Liguria & 151 & 142 & 111 -- 173\\
Lombardia & 1202 & 1202 & 1053  -- 1305\\
Marche & 127 & 120 & 94 -- 151\\
Molise & 4 & 4 & 0 -- 11\\
Piemonte & 394 & 392 & 348 -- 451\\
Puglia & 80 & 79 & 41 -- 159\\
Sardegna & 26 & 24 & 12 -- 38\\
Sicilia & 62 & 62 & 51 -- 94\\
Toscana & 256 & 240 & 203 -- 283\\
Trentino Alto Adige & 128 & 128 & 100 -- 157\\
Umbria & 39 & 37 & 28 -- 63\\
Valle d'Aosta & 16 & 18 & 9 -- 32\\
Veneto & 257 & 259 & 216 -- 299\\
\hline
\end{tabular}
\end{center}
\end{table}

\begin{figure}[!ht]
   \centering
 \begin{center}
  \includegraphics[height=8cm,width=9cm]{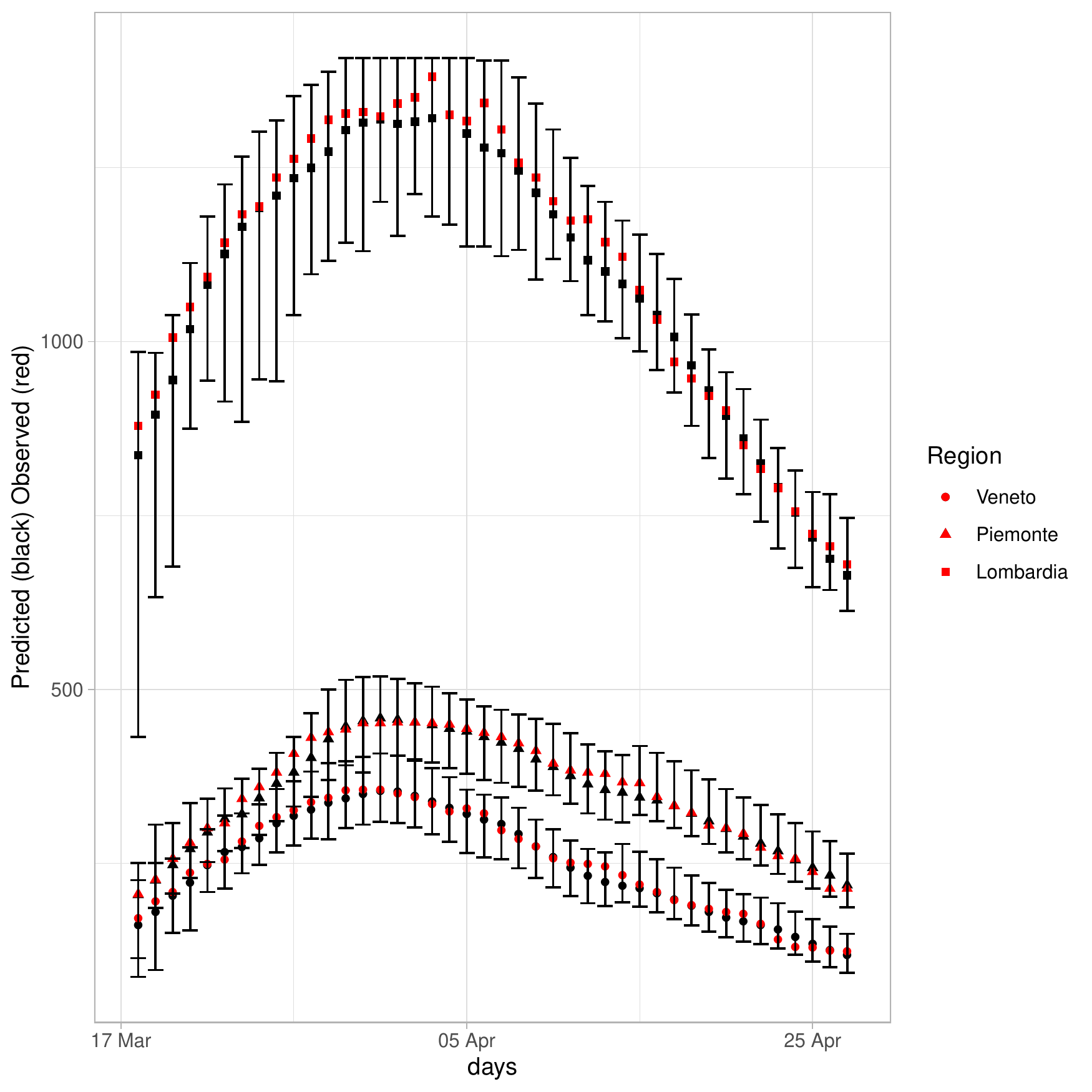}
  \end{center}
 \caption{Observed (drk) and predicted (red) values with 99\% prediction Intervals for the three northern Regions Lombardia, Piemonte and Veneto }
		\label{fig:lombPiemVen}
\end{figure}

In the following plot (Figure \ref{fig:lombPiemVen}), we provide a focus on some Regions, those whose health systems were under pressure for the longest time. The prediction intervals were slightly wide but reasonable, for instance for Lombardia, the most
severely affected Region, the median length was 20\% of the predicted occupancy. Median difference between upper limit of the prediction interval and total available beds in the
Region is 4.6\% in Veneto. We also report data and estimates for Piemonte. This is because the news overlooked the situation in Piemonte, mainly focusing on Lombardia and Veneto. However, while Veneto was more effective at containing the epidemic, the health system in Piemonte was (and partially still is) under pressure, with a decay in the ICU occupancy much slower than the other Regions.

\section{Conclusions}
\label{conclusions}

We estimated the occupancy of ICU beds at the Regional level in Italy during the Covid-19 outbreak. The resulting estimates are obtained as a combination of two approaches, which address different data features, i.e. heterogeneity and time-dependence, merged via ensambling. The proposed approach is data-driven, no simulated scenarios are required, and it is based on rather simple modelling assumptions. We have discussed in this paper only one-day ahead predictions, but our approach can also be easily extended to three-days and five-days horizons.

From the beginning of this emergency in Italy, the ICU bed capacity was identified as a major bottleneck, to be monitored to avoid the increase in the fatality rate. Accordingly, strong efforts have been dedicated to this issue from government and planners. Our approach was able to predict the demand of ICU beds since the very beginning, showing an improvement in its behavior as long as more data were available. These information and predictions were shown and freely available on a daily basis on the StatGroup-19 Facebook page (https://www.facebook.com/StatGroup-19-100907671547894). A correct communication of the evolution of the epidemic under all its variables was crucial to properly inform the general public and avoid any unmotivated concern.

As a by-product conclusion drawn from this analysis, we noticed that the Regional health systems, though under pressure, were able to properly satisfy the demand for critical care. This might be related to the restrictions imposed by the government, which effectively reduced the spread of the Covid-19 in Regions with a low number of ICU beds.


\begin{thebibliography}{}

\bibitem{Peeri2020} Peeri, N.C., Shrestha,N., Rahman, S., Zaki, R., Tan, Z., Bibi, S., Baghbanzadeh, M., Aghamohammadi, N., Zhang, W and Haque, U. (2020). The SARS, MERS and novel coronavirus (COVID-19) epidemics, the newest and biggest global health threats: what lessons have we learned? {\it International Journal of Epidemiology} dyaa033, https://doi.org/10.1093/ije/dyaa033.

\bibitem{flaxal:2020} Flaxman, S. , Mishra, S., Gandy, A., Unwin, H. J. T., Coupland, H., Mellan, T. A., Zhu, H., Berah, T., Eaton, J. W., Guzman, P. N. P., Schmit, N., Callizo, L., Imperial College COVID-19 Response Team, Whittaker, C., Winskill, P., Xi, X., Ghani, A., Donnelly, C. A., Riley, S., Okell, L. C., Vollmer, M. A. C., Ferguson, N. M.,  and Bhatt, S. (2020).  Estimating the number of infections and the impact of non-pharmaceutical interventions on COVID-19 in European countries: technical description update.  {\it arXiv:2004.11342v1 [stat.AP]}.

\bibitem{remuzzi2020} Remuzzi, A. and Remuzzi G. (2020). COVID-19 and Italy: what next? {\it Lancet}, {\bf 395}, 1225--1228.

\bibitem{diekal:2013} Diekmann, O., Heesterbeek, H. and Britton, T. (2013).  \textit{Mathematical Tools for Understanding Infectious Disease Dynamics}. Princeton University Press, Princeton.

\bibitem{grasal:2020} Grasselli, G., Zangrillo, A., Zanella, A., Antonelli, M., Cabrini, L., Castelli, A., Cereda, D., Coluccello, A., Foti, G., Fumagalli, R., Iotti, G., Latronico, N., Lorini, L., Merler, S., Natalini, G., Piatti, A., Ranieri, M. V., Scandroglio, A. M., Storti, E., Cecconi, M., Pesenti, A., for the COVID-19 Lombardy ICU Network (2020). Baseline Characteristics and Outcomes of 1591 Patients Infected With SARS-CoV-2 Admitted to ICUs of the Lombardy Region, Italy. {\it JAMA}, {\bf323}(16): 1574--1581. doi:10.1001/jama.2020.5394.

\bibitem{white2020} White, D.B. and Lo, B. (2020). A framework for rationing ventilators and critical care beds during the COVID-19 pandemic. {\it JAMA}. Published online March 27, 2020. doi:10.1001/jama.2020.5046

\bibitem{grasselli2020b} Grasselli, G., Pesenti, A., Cecconi, M. (2020) Critical Care Utilization for the COVID-19 Outbreak in Lombardy, Italy: Early Experience and Forecast During an Emergency Response. {\it JAMA}, {\bf 323}: 1545--1546.

\bibitem{sebastiani2020} Sebastiani, G., Massa, M. and Riboli, E. Covid-19 epidemic in Italy: evolution, projections and impact of government measures. {\it European Journal of Epidemiology} {\bf 35}: 341–345.

\bibitem{Giordano2020} Giordano, G., Blanchini, F., Bruno, R., Colaneri, P., Di Filippo, A., Di Matteo, A. and Colaneri, M. (2020). Modelling the COVID-19 epidemic and implementation of population-wide interventions in Italy. {\it Nature Medicine}, to appear,DOI:10.1038/s41591-020-0883-7.

\bibitem{Bohning2020} B\"ohning, D., Rocchetti, I., Maruotti, A. and Holling, H. (2020). Estimating the undetected infections in the Covid-19 outbreak by harnessing capture-recapture methods. {\it medRxiv 2020.04.20.20072629}

\bibitem{Yue2020} Yue, M., Clapham, H. E., Cook, A. R. (2020). Estimating the Size of a COVID-19 Epidemic from Surveillance Systems. {\it Epidemiology}, April 10, 2020 - Volume Publish Ahead of Print - Issue - doi: 10.1097/EDE.0000000000001202

\bibitem{Chen2020} Chen, Y.C., Lu, P.E., Chang, C.S. (2020) A Time-dependent SIR model for COVID-19. arXiv preprint arXiv:2003.00122, 2020.

\bibitem{breslow1993} Breslow, N. E. and Clayton, D. G. (1993). Approximate inference in generalized linear mixed models. {\it Journal of the American Statistical Association}, {\bf 88}, 9--25.

\bibitem{zhang2011} Zhang, H., Lu, N., Feng, C., Thurston, S. W.,  Xia, Y., Zhu, L.  and Tu, X. M.  (2011). On fitting generalized linear mixed-effects models for binary responses using different statistical packages. {\it Statistics in Medicine}, {\bf 30}, 2562--2572.

\bibitem{kim2013} Kim, Y., Choi, Y.-K. and Emery, S. (2013). Logistic regression with multiple random effects: a simulation study of estimation methods and statistical packages. {\it The American Statistician}, {\bf 67}, 171--182.

\bibitem{fokianos2009} Fokianos, K., Rahbek, A. and Tjøstheim, D. (2009). Poisson autoregression. {\it Journal of the American Statistical Association}, {\bf 104}, 1430--1439.

\bibitem{agosto2016} Agosto, A., Cavaliere, G., Kristensen, D. and Rahbek, A. (2016). Modeling corporate defaults: Poisson autoregressions with exogenous covariates (parx).  \textit{Journal of Empirical Finance} \textbf{38}, 640--663.

\bibitem{chen2016} Chen, C. W. and Lee, S. (2016). Generalized poisson autoregressive models for time series of counts. {\it Computational Statistics \& Data Analysis}, \textbf{99}, 51--67.

\bibitem{fokianos2011} Fokianos, K. (2011). Some Recent Progress in Count Time Series. {\it Statistics}, {\bf 45}, 49--58.

\bibitem{bollerslev1986} Bollerslev, T. (1986). Generalized autoregressive conditional heteroskedasticity. {\it Journal of Econometrics}, {\bf 31}, 307--327.

\bibitem{tjostheim2012} Tjøstheim, D. (2012). Some recent theory for autoregressive count time series. {\it Test}, {\bf 21}, 413--438.

\end{thebibliography}
\end{document}